# Semantic Data Warehouse Modelling for Trajectories

Michael Mireku Kwakye [A, B]

[A] *Faculty of Computing and Information Systems, Ghana Technology University College, PMB 100, Accra, Ghana, mmireku@gtuc.edu.gh*

[B] *Department of Computer Science, University of Calgary, 2500 University Drive NW, Calgary, Alberta, Canada, michael.mirekukwakye@ucalgary.ca*

**ABSTRACT**

*The trajectory patterns of a moving object in a spatio-temporal domain offers varied information in terms of the management of the data generated from the movement. A trajectory data warehouse is a data repository for the data and information of trajectory objects and their associated spatial objects for defined temporal periods. The query results of trajectory objects from the data warehouse are usually not enough to answer certain trend behaviours and meaningful inferences without the associated semantic information of the trajectory object or the geospatial environment within a specified purpose or context. This paper formulates and designs a generic ontology modelling framework that serves as the background model platform for the design of a semantic data warehouse for trajectories. This semantic trajectory data warehouse can be adaptable for trajectory data processing and analytics on any chosen spatio-temporal application domain. The methodology underpins on higher granularity of data as a result of pre-processed and transformed ETL data so as to offer efficient semantic inference to the underlying trajectory data. Moreover, the approach outlines the thematic dimensions that serve as necessary entities for extracting semantic information. Additionally, the modelling approach offers a design platform for effective predictive trend analysis and knowledge discovery in the trajectory dynamics and data processing for moving objects.*

**INTRODUCTION**

The trajectory of an object is the sequence of ordered-points of a path or route followed by the moving object in a defined geographical space, and mostly within a specified temporal function. Objects could be classified as humans, animals, or vehicles within a geographic context. The trajectories of these objects could be affected or informed by varied factors, such as, weather changes (for example, wind resistance, rain precipitation, and snow falls), gravity, geographic surfaces, and events, amongst others.

Over the years, the tracking of the trajectory movement of these objects has been done with the aid of devices, such as, Global Positioning Systems (GPSs), smart phones, geo-sensors, surveillance cameras, and Radio-Frequency Identification (RFID) tags, amongst others. The trajectory data captured by these devices and the subsequent processing by Geographical Information Systems (GIS) have increased the amount of movement data for future trend analysis (Rigaux *et al.*, 2001, pp. 2). Moreover, recent application of satellite devices has been adopted in capturing the trajectory of moving objects, as has been in the case of vehicular movements (Bodur & Mehrolhassani, 2015).

The concept of trajectories has received quite an appreciable amount of study in the literature. These studies have gained much attention because of the traditional spatial and temporal applications in the fields of GIS. The information gathered from specific trajectories has become useful in varied application domains. The collected data are valuable in the detection of informed and uninformed trends in trajectory movements, and critical in decision-making for these application domains, such as, tourism management, and animal migration, amongst others.

A semantic trajectory data warehouse (DW) is a data repository that stores the semantic information of a trajectory object and the associated spatial objects within a period of temporal instance to achieve a particular goal or purpose. In analyzing the semantic information of a trajectory object, the stop and move activity of the object at any temporal instance is determined by the goal. This goal could be classified as being a personal goal, an activity object goal, or a movement objective, for the activity in view (Da Silva *et al.*, 2015; Parent *et al.*, 2013). Some of the semantic information are annotations of semantic objects, events and activities, behavioral data, and the data of the spatio-temporal objects, amongst others.

In this paper, a framework model for the modelling and design of a semantic data warehouse for a trajectory object in a spatio-temporal paradigm is formulated. The novel proposition is to gather all relevant semantic data of objects and events related to the trajectory object, as well as, the trajectory object itself and its movement dynamics. Hence, the main contribution in this paper is to present a generic modelling approach for the design of semantic data warehouses for trajectories. To this end, an approach based on the higher granularity level of preprocessed and transformed trajectory data is placed in focus, being an output of Extract-Transform-Load (ETL) procedures. It is of the expectation that this higher granularity level of trajectory data will offer an optimal medium for efficient query processing, and to accommodate the large volume of trajectory data feeds from raw sources.

The data gathered from raw trajectory data from different application domains keep rising and have become much more relevant because of the varied inferences that can be drawn from them. Much more is the need to find out and ascertain why some trajectory objects behave, move, or stop at specific areas and at specific times in their geospatial environment.

Modelling and designing a trajectory data warehouse for analysis and predictive inferences has been studied so far in the literature, and have been addressed by varied researchers. On the other hand, the need to understand the rare semantics of trajectory stops, moves, velocity rates, and movement pattern, amongst others, still remains a challenge. Additionally, the characteristic attributes of trajectory objects and the semantic annotations associated with events and activities that the objects participates in draw out vital information that most trajectory data warehouses are not able to address.

Past research on trajectory data warehousing have focused on studying the modelling and design constructs for the warehouse structure. Some of the methodology approaches have formulated ontologies to utilize them as the modelling framework constructs. Most of these methodology approaches do not outline a comprehensive generic ontology model for the trajectory data warehouse, and the associated semantic annotations. Moreover, some of the approaches do not consider inferences that can be drawn from social media interactions of the trajectory object.

In this research, a novel methodology approach that defines and outlines a generic ontology model for handling the varied semantic characteristics of trajectory objects, events and activities, environmental considerations, as well as, social media interaction is introduced. This generic ontology model serves as the background platform for the modelling and design of the thematic constructs for a semantic trajectory data warehouse. The formulated semantic trajectory data warehouse offers a data repository platform for detailed and enhanced fact attributes and numeric measures. Additionally, the sound generic ontology model enables the semantic trajectory data warehouse to offer descriptive dimensionality attribute representation for the unique characteristic of any chosen application domain.

More specifically, a generic semantic trajectory data warehouse that can be related to varied application domains is addressed, even in the face of peculiar characteristic features. The merits of the methodology approach offers:
1. An expressive generic ontology model for trajectory objects, geographic environments, events and activities, and social media interaction;
2. A comprehensive trajectory data warehouse platform for efficient, scalable, and optimized query processing;
3. Maximum semantic annotation enrichment for every aspect of the trajectory of a moving object.

**Definition 1.** **(Certain Query):** A *Query*, $Q$ is said to be *Certain* for all Instances, $\mathcal{I}$ and Properties, $\mathcal{P}$ of a Multidimensional Database, $\mathcal{MD}$ iff $Q \vDash \mathcal{I}$, such that $\mathcal{I} \subseteq \mathcal{MD}$ and $Q$ satisfies $\mathcal{P} \in \mathcal{MD}$ ∎

**Definition 2.** **(Certain Answer):** A *Tuple*, $\mathcal{T}$ forming an *Answer* to a certain query, $Q$ is said to be *Certain* iff $\mathcal{T} \vDash Q$ for all Instances, $\mathcal{I}$ of Multidimensional Database, $\mathcal{MD}$ and $\mathcal{T}$ fulfils $\mathcal{I} \in \mathcal{MD}$ ∎

A summary of some assumptions needed to validate this research is enumerated, as follows: First, the use of high granularity preprocessed ETL instance data is adopted for population into the designed semantic trajectory data warehouse. This is necessary because of the highly-refined and aggregated data item elements in the fact or dimension repositories. Second, the processing of *certain queries* on the formulated semantic trajectory data warehouse. *Certain query answers* are expected because of the distinct definition of fact attribute measures and dimension attributes in the trajectory data warehouse.

The technical contributions are summarized, as follows;
- To formulate a generic ontology for the modelling of semantic trajectory of moving objects which extends to different application domains;
- To instantiate the constructs of the formulated generic ontologies to design a semantic data warehouse model for the trajectory data of moving objects;
- To outline the thematic dimensions, the fact information, and the attribute and measure data for the trajectory data warehouse; which will serve as data modelling entities for semantic trajectory of moving objects;
- To utilize the semantic data warehouse instances as a platform for the predictive trend analysis and knowledge discovery of the trajectory of moving objects in a spatio-temporal application domain; such as, tourist movement and tourism management, birds migration, and traffic management, amongst others.

The rest of the paper organized, as follows: In Section 2, the background studies that have been conducted so far in the literature relating to spatio-temporal and trajectory data warehouse is addressed and reviewed. Section 3 addresses an overview of the trajectories, semantics, and semantic trajectory data warehouse which serves as a modelling design platform for predictive analysis and decision-making. The thematic object representation of the modelling design is presented in Section 4, where the detailed modelling constructs and their semantic relevance is described. Section 5 discusses the experimental implementation procedures of ontology modelling, ETL procedures, as well as, design work of the trajectory data warehouse.

In Section 6, some application domain areas regarding this research is discussed. Moreover, a summary of the key outcomes arising out of this research is discussed; and there is a highlight on some of the prime queries that can be posed to the semantic trajectory data warehouse. Section 7 discusses the comparative analysis and performance measurements of previous related work to the propositions in this research. Here, the merits of the proposed research approach over the other methodologies is addressed. Section 8 concludes and discusses the major contributions, and also highlight on the open issues and future work in the area of semantic trajectory data warehouse modelling.

## BACKGROUND RELATED WORK

A number of related studies have been investigated in the areas of spatio-temporal data warehousing and trajectories. These studies in the literature tend to serve as knowledge base for main contribution in this paper, and offer a broad platform to present the modelling design. One of the earliest propositions of conceptual modelling of spatio-temporal applications was investigated by Parent *et al.* (2006). In their study, the authors highlighted the data modelling from a multidimensional view where each dimension is handled from an orthogonal theme of data structures, space, and time representation, amongst others.

## Trajectories and Data Warehouses

Studies on trajectory data warehouses (DWs) have been investigated by Orlando *et al.* (2007), and Vaisman and Zimányi (2013), with the latter being a more recent study. The work by Orlando *et al.* (2007) investigated and addressed the challenges of aggregations that are encountered as a result of building trajectory DWs. In their assessment, the authors propose a methodology to outline the complex aggregate and summarization computation of measure presence. Vaisman and Zimányi (2013), on the other hand, presented a textual description of the constituent dimensions of a data warehouse for trajectories, where the authors discussed various mobility data analysis, varied temporal types, and instances of queries that could be posed to trajectory data warehouses. Their work finally analyzed a typical instance of *Northwind* trajectory data warehouse to support their conceptual proposition.

Spaccapietra *et al.* (2008), also in their study of the conceptual approach of trajectories propose a modelling methodology. On one hand, their approach adopted a set of standard constructs that enriches the underlying spatio-temporal data model. On the other hand, customized constructs are adopted to offer maximum flexibility for specific semantics of application-centered trajectories. The two approaches of modelling that were stipulated are based on data types and on design patterns. In presenting their proposition, the authors make mention of semantics of Point-of-Interests (POIs) object, but fail to give a detailed description and impact analysis of the semantics of these POIs object, the trajectory object, and also the movement dynamics of the trajectory object.

## Trajectories and Ontologies

An ontological approach for semantic representation of trajectories was investigated by Baglioni *et al.* (2009). The authors applied procedures of semantic enrichment of trajectories by way of deducing reasoning from the trajectory patterns as a result of mining the raw data feeds. This study brought to the fore a background knowledge on which the authors in Campora *et al.* (2011) presented a closer study of the modelling of data warehouses for trajectories. The work by Campora et al. (2011) leveraged on and complemented the other prior research work investigated in Marketos *et al.* (2008) and Spaccapietra *et al.* (2008). Thus, Campora *et al.* (2011) introduced formal modelling constructs for high-level expression of the architecture and modelling of trajectory data warehouses.

Parent *et al.* (2013) addressed the broad overview of data analysis of trajectories and mobility data management based on the studies done in the literature so far. In their survey, the authors expounded on the approaches and techniques of trajectory construction, the enrichment of semantic information on trajectories, and the application of data mining techniques to analyze and extract semantic knowledge from trajectory movement.

## Recent Approaches and Notable Propositions

Recent studies on the semantic modelling of trajectory data warehouses have been investigated by Wagner *et al.* (2014), Sakouhi *et al.* (2014), Da Silva *et al.* (2015) and Manaa and Akaichi (2016). I address an overview of the key propositions and contributions of the studies performed.

### *SWOT: Conceptual Data Warehouse Model for Semantic Trajectories*

More specifically, the work by Da Silva *et al.* (2015) presented a study on formulating a conceptual and semantic data warehouse for trajectories where they proposed a model that relies on the DOGMA framework (Jarrar & Meersman, 2008) and offers a dual modelling of ontologies. This proposition enables the separation of the ontology information into two conceptual data layers. The ontologies are, namely; *Consensual* and *Interpretation*. Consequently, the authors fail to address the addition of much enhanced numerical measures for the fact relationship and did not illustrate a generic ontology model which can be applicable to different domains.

*Ontology-Based Trajectory Data Warehouse Conceptual Model*

In the study by Manaa and Akaichi (2016), the authors discussed the approach of modelling ontology data using Ontology-Based Moving Object Data (OBMOD); and the efficient ways of storing and querying heterogeneous OBMOD. In their methodology, the authors defined an ontology-based design approach to model and analyze a global trajectory shared ontology and its associated semantics. Moreover, their approach defined the structure of the conceptual model for a semantic trajectory data warehouse based on a formulated algorithm, but fail to outline practical query processing on the semantic trajectory data warehouse.

In the review of the literatures, it can be inferred that though semantic trajectory data warehouse has been studied so far, the proposition for a generic semantic trajectory data warehouse model for varied application domains has not been yet addressed comprehensively. Most importantly, the ability of the generic semantic trajectory data warehouse to incorporate social media interaction of the trajectory object is still unavailable.

In this paper, the research leverages on the prior work by Da Silva *et al.* (2015) and Manaa and Akaichi (2016), and proposes a methodology that defines a complete generic ontology framework. Here, this ontology framework will globally analyze a trajectory data and the associated semantics for the trajectory objects, the spatio-temporal object dynamics, and the events representation in the trajectory movement. A novel contribution to the proposed ontology design is to analyze the social media interaction of a trajectory object, such as, a tourist posts and comments as he or she travels in a trajectory path. Additionally, the proposed methodology framework will define a formulated multidimensional star-schema model for the semantic trajectory data warehouse based on the earlier proposed generic ontology framework.

Moreover, a key contribution in this research is to deliver a data repository that is able to answer practical queries for semantic analysis of trajectory data based on enhanced numeric measures and descriptive dimension attributes. The proposition of the modelling and design of a trajectory data warehouse is based on the intuition of a higher level ETL-transformed trajectory data.

## TRAJECTORY ANALYSIS, SEMANTICS AND DATA WAREHOUSING

The trajectory of a moving object is usually focused on the path movement of the object and the relationship to the geographic locations that the object interacts with, and geometric properties like surface area, velocity, and direction change associated with the movement. This Section discusses and explain the conceptual idea regarding trajectory and mobility, and the subsequent modelling of their constituent data for data warehousing purposes.

### Trajectory Analysis

The study of the trajectories involves the identification of *stops* and *moves* within the path of the trajectory. A *stop* is a non-empty time interval of which the trajectory object does not move (Spaccapietra *et al.*, 2008). Each *stop* could be prompted or activated by the need or involvement in an activity of the trajectory object at that specific temporal instance, and as well as, its association to a geographic object. An example of a *stop* is a bird resting in a nest on a mountain top. The *move* is a part of a trajectory representing a spatial range that is delimited by two distinct *stops*. As a result, a *move* is a set of time-varying points defined from two consecutive *stops*, that is, $\{T_{Begin}, T_{FirstStop}\}$ or $\{T_{LastStop}, T_{End}\}$ (Spaccapietra *et al.*, 2008). In context, the set of trajectory *stops* and *moves* together with the temporal instances, in the overall path movement of an object generates a *raw trajectory* for the moving object.

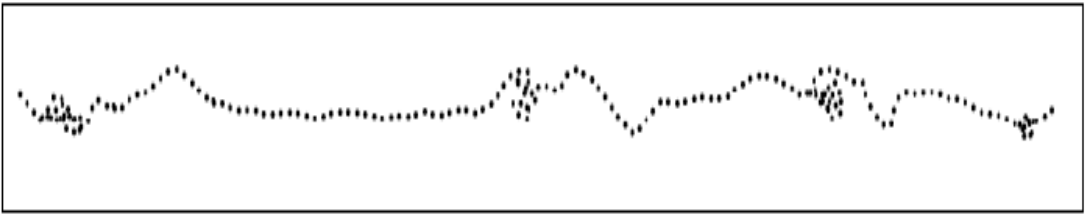

*Figure 1. An Illustration of a Raw Trajectory*

**Definition 3. (Raw Trajectory):** *Let $G = \{p_1, p_2, \ldots, p_n\}$ represent the arbitrary set of noticeable points in a geographical space, G. Let each point be identified as, $p_i = (x_i, y_i)$, where $x_i$ and $y_i$ are the geographic latitude and longitude coordinates, respectively. A raw trajectory, T, is an ordered list $T: \{\exists G\,((p_1, t_1), (p_2, t_2), \ldots (p_m, t_m))\}$, where each $t_i$, for $i = (0, 1, \ldots, m)$, is the timestamp at which the trajectory object was stationed, and $t_1 < t_2 < t_3 < \cdots < t_m$* ∎

*Figure 1* illustrates a representation of a raw trajectory for a moving object. Here, various stops, with time stamps, have been identified whereas stops that do not have any relevance to the application domain have been eliminated. Moreover, there are certain segments of the entire trajectory where trajectory points are dense. The densely pointed trajectories indicate a possible semantic activity for the trajectory object's movement. These semantic activity points give rise to the concept of semantic annotation or semantic segmentation of trajectories.

Each trajectory of a moving object is based on an overall goal or purpose of the movement. For example, during each Fall (Autumn) season, birds migrate annually from Europe (Northern Hemisphere) to Africa (Southern Hemisphere) in search of better food availability, and do a reverse migration in the Spring season back to Europe for breeding. This overall set of goals or purpose of the trajectory movement informs the various *stops* and *moves* for the trajectory movement. As a result, each *stop* has a unique goal or purpose which can vary even among different entities in a single group of trajectory objects (Moreno *et al.*, 2010).

## Semantic Enrichment for Trajectories

The study and analysis of trajectories have provided much knowledge in the movement dynamics of objects, such as, in bird migration, and tourist movement, amongst others. Whereas these trajectory objects are prompted to varying *stops* and *moves* at particular geographic objects and certain temporal instances, the need to study and understand the meaning or reason motivating the goal of *stops* and *moves* have become evident (Moreno *et al.*, 2010; Spaccapietra *et al.*, 2008). The meaning attached to a trajectory *stop* or *move* is usually formalized as a semantic annotation that offers a contextual information to the analysis of trajectory patterns.

**Definition 4. (Semantic Annotation):** *Let $T = \{T_1, T_2, \ldots, T_n\}$ represent a set of raw trajectory points, for $0 \leq i < n$. Let the trajectory segments, S represent a disjoint or overlapping set of trajectory points, such that, $S = \{S_1, S_2, \ldots, S_m\}$ and $S_1 = \{T_1, T_2, \ldots, T_k\}$, $S_2 = \{T_{k+1}, T_{k+2}, \ldots, T_n\}, \ldots, S_z = \{T_n, T_{n+1}, T_{n+2}, \ldots, T_e\}$, where $T_1$ and $T_e$ are the first and last trajectory points, respectively, and for $k > 0, n > 0, e > 0$, and $k < n < e$. A Semantic Annotation, $S_A$ is contextual domain information that is associated with each trajectory segment, $\{S_1, S_2, \ldots, S_m\}$, in which we have a pair $< S_1, S_A >, < S_2, S_A >, \ldots, < S_m, S_A >$, such that each, $S_A = \{O_P, O_G, O_E\}$; where $O_P$ is the Geographic Object Property, $O_G$ is the Trajectory Object Goal, and $O_E$ is the Event that the Trajectory Object participates in.* ∎

Semantic enrichment for trajectories tends to classify the meaning and what informs a trajectory object to *stop* or *move* at a point based on varied factors. Some of these factors could be identified as; the properties of the geographic object or POIs that initiate the *stop* (for e.g., a community festival occurring at certain times of the year on a segment of the road), the weather and environmental factors present (for e.g., the rain precipitation could be high to prompt a bird to rest), the activities occurring at a POI during a temporal instance (for e.g., there could be an arts exhibition at a museum), and the mode of transportation needed to move in between *stops* (for e.g., cars can only move on roads and at certain speeds at some parts of the road), amongst others.

## Trajectory Data Warehousing

The design and modelling of a data warehouse for the trajectories offers a platform for query processing and predictive trend analysis for the trajectory object. The main object in constructing a data warehouse is to be able to collect, clean, and streamline the raw trajectory data over a period of time into a permanent data repository.

Building a trajectory data warehouse involves defining the various dimensions, such as, geographical space, temporal instance, trajectory object instance, as well as, events associated with each point of the entire trajectory of the moving object. A fact data repository is formulated to store the numeric measures associated with each trajectory point or segment, and also establish a referential relationship to each of the dimensions. Moreover, the data sources covering these dimensions are identified and prepared (cleaned) for subsequent procedures. To populate the trajectory data warehouse with instance data, basic Extract-Transform-Load (ETL) procedures are adopted in processing and transforming the raw trajectory data into the dimensions defined in the data warehouse.

Various approaches in modelling and designing a trajectory data warehouse have been proposed in the literature, and as discussed in the background related work in Section 2. The next Section 4 discusses indepthly the novel approach of a generic semantic trajectory data warehouse, which can be related to varied application domains.

## GENERIC SEMANTIC TRAJECTORY DATA WAREHOUSE – THEMATIC CONSTRUCTS

This Section describes in detail the modelling proposition for semantic trajectory data warehouse and present the conceptual intuition that motivate this research approach; whilst expounding on the thematic objects of the modelling design. To this end, instance examples to some of the objects are provided to illustrate a practical way of modelling these objects.

The proposition for the approach of modelling and design is based on the adoption of the orthogonal methodology, as addressed by Parent *et al.* (2006). The author therefore models the design based on the conventional Multidimensional Entity Relationship (MER) notation for spatio-temporal data warehouse modelling (Parent *et al.*, 2006). The data modelling approach for the multidimensional views of the metadata of trajectories, as well as the instance data, offers the ability to mine and detect informed and uninformed trends in the trajectory data.

The modelling design, in star-schema data warehousing model, is represented in 5 main dimensional themes for a single fact table. A star-schema model for the data warehouse ontology is chosen because of its simplest representation of analytical data. Moreover, the characteristic feature of hierarchical aggregation and summarizations of attribute data in the dimensions are easily referenced to the fact attribute and measure data (Sapia *et al.*, 1998; Tryfona *et al.*, 1999). The modelling ontology is fashioned as consisting a fact data store with an *n-ary* relationship to each of the dimensional themes.

The fact data store additionally keeps a two main measure information, namely, enhanced numeric and statistical aggregation measures. The enhanced numeric measures are instantiated as, *square area*, *overall temporal duration*, *number of semantic stops*, and the *number of mobility modes*, amongst others (see *Figure 2*). The statistical aggregation measure data are instantiated as, *average trajectory speed*, *average even time duration*, *minimum activity duration per event*, and the *maximum trajectory travel distance*, amongst others (see *Figure 2*).

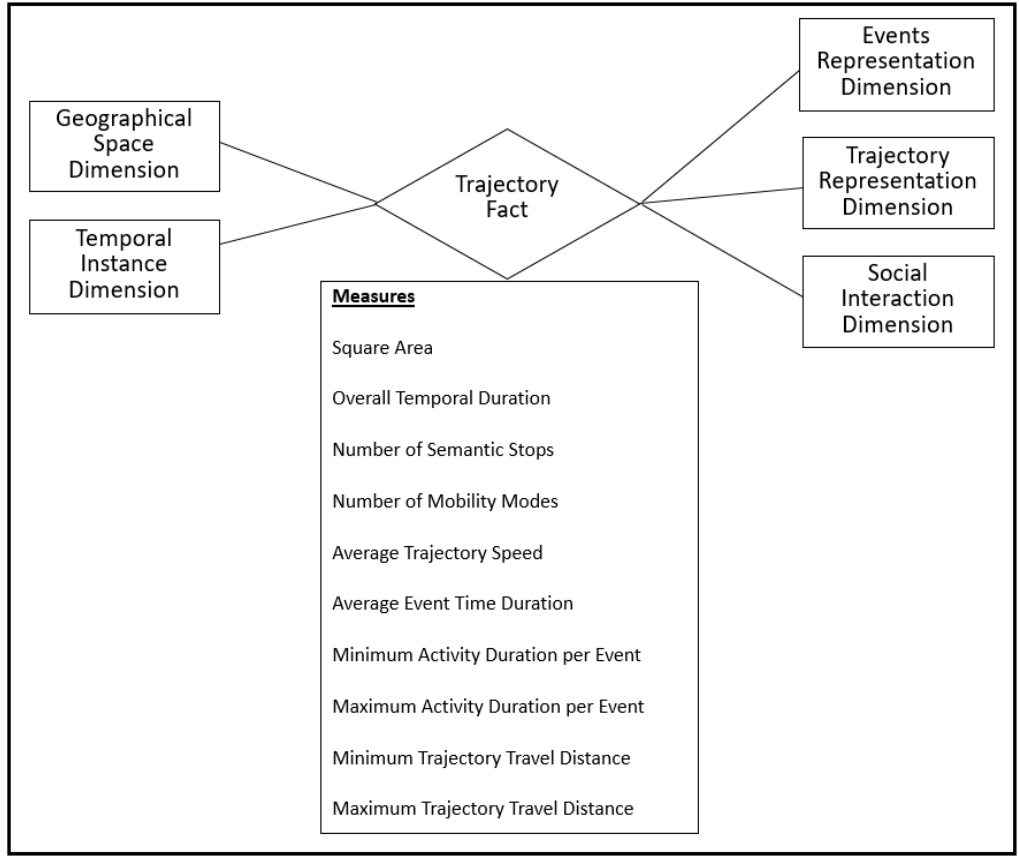

*Figure 2. Conceptual Semantic Trajectory Data Warehouse Model*

A graphical representation of the generic model of the semantic trajectory data warehouse is displayed in *Figure 2*. In the illustration, the dimensional themes outlined are *Geographical Space*, *Temporal Instance*, *Events Representation*, *Trajectory Representation*, and *Social Interaction*. An overview of the characteristics and information regarding these dimensions are addressed, and the constructs that inform each of these dimensional themes in the subsections that follow, are explained.

## Geographical Space Dimension

The Geographical Space thematic dimension represents the spatial extent for the model design. As trajectory objects move in a geographical space, the need to extract semantic meaning from these objects that they associate becomes necessary. A typical geographical space dimension is composed of the *Continent*, *Country*, *State* or *Province*, *Region*, *City*, and *District*. It will be noted that these hierarchical levels form the background aggregation for additional level representation in the specification of this dimension.

**Definition 5. (Point-of-Interest):** *Let $G$ represent a geographic space for a trajectory object. A Point-of-Interest, $P = <G, O, S_S, S_M>$ is a quadruple (4-tuple) consisting of a Semantic Stop, $S_S$ and/or Semantic Move, $S_M$ at a geographic object, $O$ where the trajectory object visited in its trajectory movement.* ∎

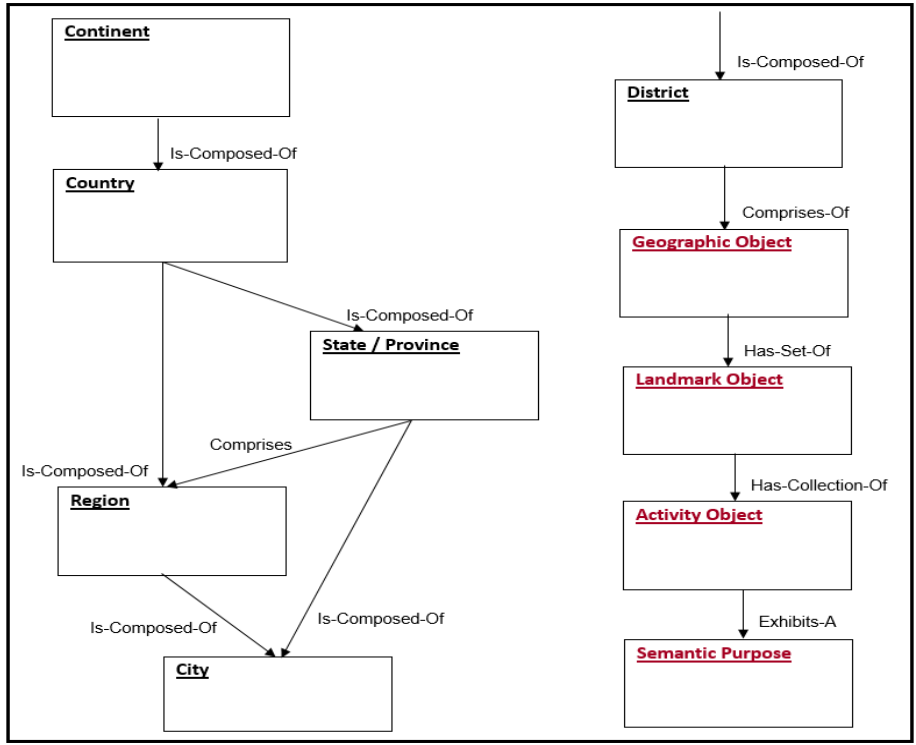

*Figure 3. Taxonomy Model of the Geographical Space Dimension*

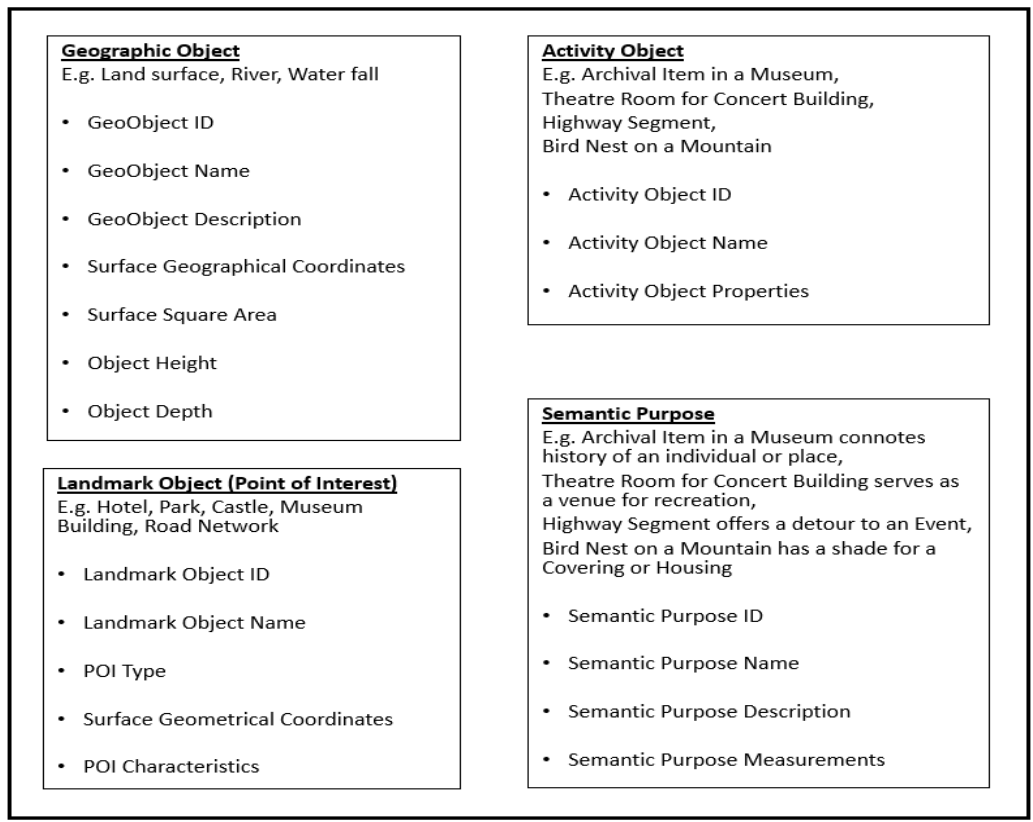

*Figure 4.    Characteristic Description of Geographical Space Taxonomy Model*

A representation of other levels depicts 3 key hierarchical levels (highlighted in red font colour in *Figure 3*), amongst other traditional levels, which are vital in generating semantic information. These are *Geographic Object* (for e.g., land surface, river, and water fall), *Point-of-Interest* (POI)/*Landmark Object* (for e.g., Hotel, Castle, Museum, and Highway Segment), *Activity Object* (for e.g., Archival Item in a Museum, Bird Nest on a Mountain, and Bridge on a Highway Segment), and *Semantic Purpose* (for e.g., Archival Item in a Museum connotes history of an individual or place, Bird Nest on a Mountain has a shade for a Covering or Housing). The hierarchy levels represented in this dimensional theme enable analytical procedures of aggregation and summarization of the dimensional data for the geographical space in perspective.

*Figure 3* displays an illustration of the taxonomy ontology model of the geographical space dimension that depicts the composition of all the hierarchical levels. In *Figure 4*, the illustration outlines an elaborate attribute definition and characteristic outlook of the *Geographic Object*, the *Landmark Object*, the *Activity Object*, and the *Semantic Purpose*. Here, each of these hierarchical levels are defined with feature data item elements that uniquely identifies the hierarchical level and allows its incorporation in aggregation and summarization.

## Temporal Instance Dimension

The Temporal Instance thematic dimension represents the time and date extensions for the movement of trajectory objects. The consideration and extraction of semantics are usually contextualized, and each context within any geographical space is associated with a temporal instance. This dimension further helps in trend and behavioural analyses of the object instances of *Events* and the *Trajectory Object*.

Some of the hierarchical levels in the dimension are *Year*, *Quarter*, *Season*, *Month*, *Week*, *Day*, *Hour*, *Minute*, and *Second*, amongst others. There could be various variations of any of these hierarchical levels, such as, a *Week* is composed of *WeekDay* and *WeekEndDay*. *Figure 5* displays the taxonomy ontology model of the temporal instance dimension which depicts each of the hierarchical levels for a temporal function.

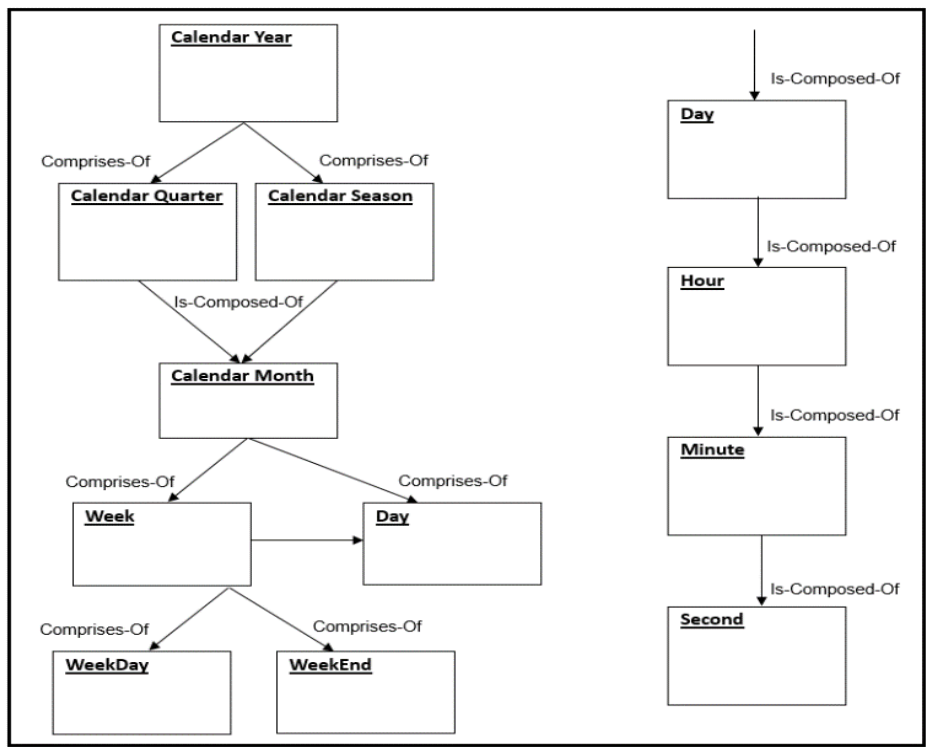

*Figure 5. Taxonomy Model of the Temporal Instance Dimension*

## Events Representation Dimension

**Definition 6. (Event-of-Interest):** *An Event-of-Interest, $\mathcal{E} = <G, E, t_i>$, is a triple (3-tuple) consisting of a geographic space, $G$, the occurance of an event of situation in the geographic space, $E$, and during a temporal period function, $t_i$, for $i > 0$.* ∎

The Events Representation thematic dimension represents the set of occurrences, incidents, experiences and episodes that are associated to the movement of a trajectory object. These events define or reveal the reason or purpose for which a trajectory object will move in a certain direction, location, or velocity rate. Moreover, the behavioural expression of a trajectory object is depicted by this dimension.

A representation of this dimension is modelled in 4 main hierarchical levels. These are; namely, *Event Item* (for e.g., Theatre Concert, Museum Exhibition, Bridge Repair, and Bird Feeds), *Goal* (for e.g., Entertainment at Concert, Scientific Interest at Museum), *Activity* (for e.g., List of Actions at Concert, Set of Stages for a Bird To Feed, and Procedures Steps at a Museum Exhibition), *Environmental Information* (for e.g., Snow Effects, Rain Precipitation, Temperature, and Wind Pressure).

*Figure 6* displays the taxonomy ontology model of the events representation dimension depicting all the hierarchical levels, as well. In *Figure 7*, the characteristic description for each of the hierarchical levels in the model is illustrated. Each level is depicted with a set of attribute data item elements that will clearly define all the features for each event associated with the trajectory dynamics of a moving object in a geospatial environment at any temporal instance.

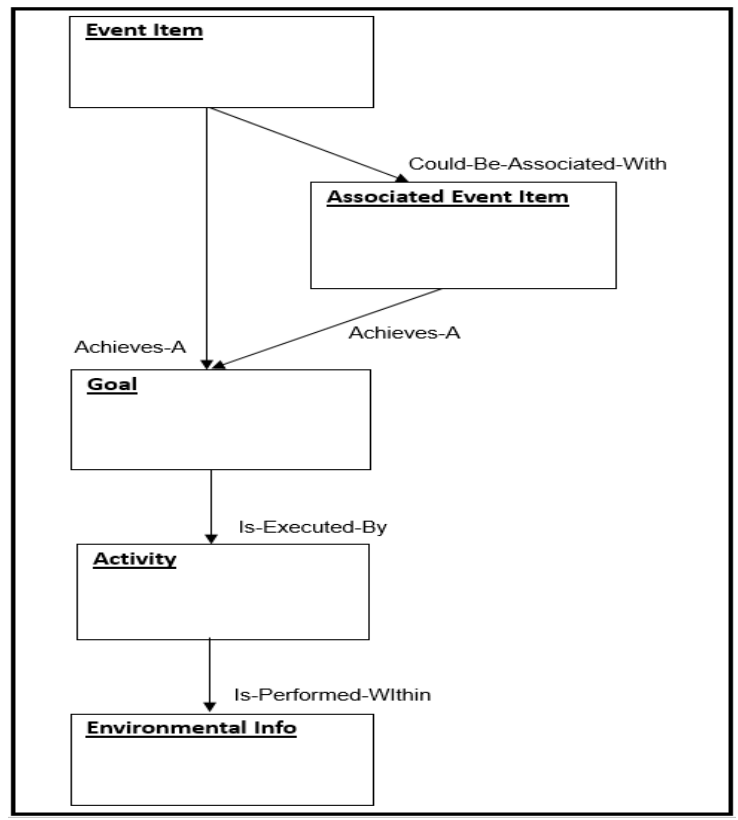

*Figure 6.   Taxonomy Model of the Events Representation Dimension*

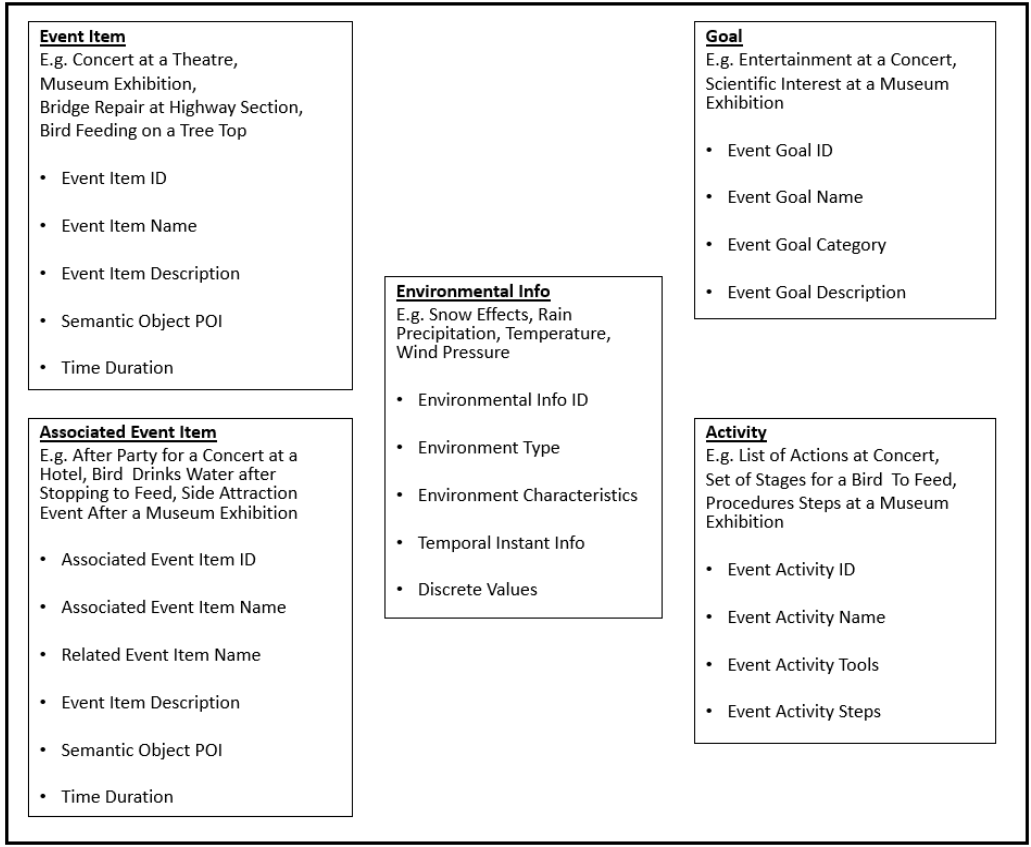

*Figure 7. Characteristic Description of the Events Representation Taxonomy Model*

**Trajectory Representation Dimension**

The Trajectory Representation thematic dimension represents features, characteristics, behavioural moods, and transportation mode that a trajectory object exhibits as part of its movement in a geographical space. Each trajectory object or a collection of objects move in a spatio-temporal instance based on a particular event(s), for a goal purpose, and execute a set of activities during the sequences of *stops* and *moves* in the trajectory.

**Definition 7. (Semantic Trajectory):** *Let $T = \{T_1, T_2, \ldots, T_n\}$ represent a set of raw trajectory points, such that each point, $T_i = \{x_i, y_i, t_i\}$, for $0 \leq i < n$. Let each trajectory segment be an ordered list, $S_k = (S_B, S_E, S_M, S_S, S_A, S_T, S_G)$, where $S_B = \{T_i, PointOfInterest\}$, for $i \geq 0$ and $S_E = \{T_j, PointOfInterest\}$, for $i < j \leq n$, $S_M$ is a Semantic Move, $S_S$ is a Semantic Stop, $S_A$ is a Semantic Object Activity, $S_T$ is a Semantic Transportation Mode, and $S_G$ is a Semantic Goal. A semantic trajectory, ST, is a finite sequential set of segments, such that, $ST = \{S_k, S_{K+1}, \ldots, S_m\}$, for $0 \leq k < m$, where each segment, $S_k$, is associated with a Semantic Annotation about an event or the moving object and its activity goal.* ∎

The goals, set of activities, and the semantic inferences and annotations together enrich a raw trajectory to form a semantic trajectory, of which its feature characteristics are captured in this dimension. *Figure 8* illustrates the semantic trajectory movements of an object, together with the semantic annotation that are associated with each *stop* or *move* of the object in the geographical space. In the diagram, 3 different transportation means are observed; by metro transit bus, by foot, and by bicycle. Additionally, there are different trajectory goals, such as, going to the central park, taking pictures, and having lunch.

A representation of the trajectory dimension is modelled by the following hierarchical levels; namely, *Trajectory Object Type* (for e.g., Human Being, Animal, etc.), *Trajectory Model* (for e.g., Tourist for Human Being, and Bird for Animal), *Trajectory Model Goal* (for e.g., Tourist visiting a Monument, and Bird Feeding), *Trajectory Model Activity* (for e.g., Set of Actions by a Tourist at Museum Exhibition, and Set of Movements for a Bird To Feed, amongst others), *Trajectory Model Behaviour* (for e.g., Velocity Rate by Tourist, Flight Velocity for Birds, and Individual or Collective Movement), *Transportation Mode* (for e.g., Air, Water, and Land, amongst others), *Transportation Type* (for e.g., Walking/Biking/Driving for Tourist, and Flight for Birds), and *Transportation Object* (for e.g., Bike for Tourist, and Soaring Flight for Birds).

The taxonomy ontology model of the trajectory representation dimension, where each of the hierarchical levels is uniquely characterized is displayed in *Figure 9*. *Figures 10* and *11* illustrate an elaborate definition and characteristic outlook of the trajectory representation dimension. In both diagrams, the attribute data item elements that define each of the hierarchical levels in the Trajectory dimension are outlined.

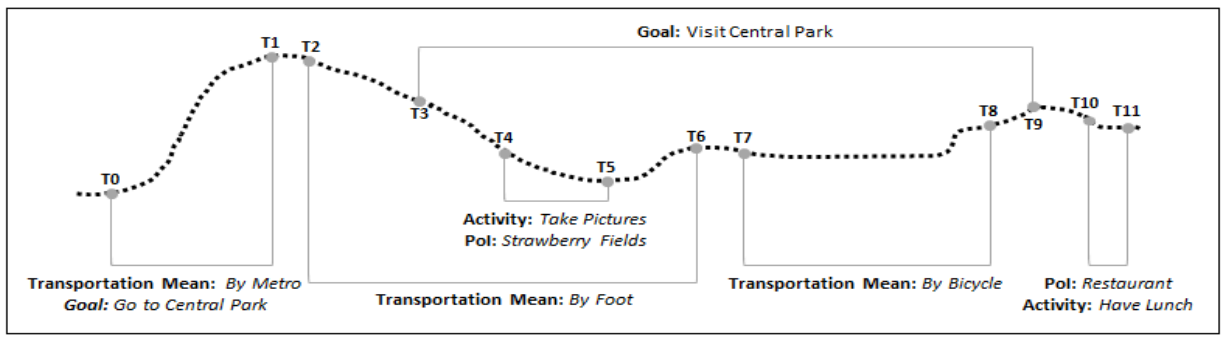

*Figure 8.*      *An Illustration of a Semantic Trajectory (Da Silva et al., 2015)*
Source: Da Silva et al., 2015

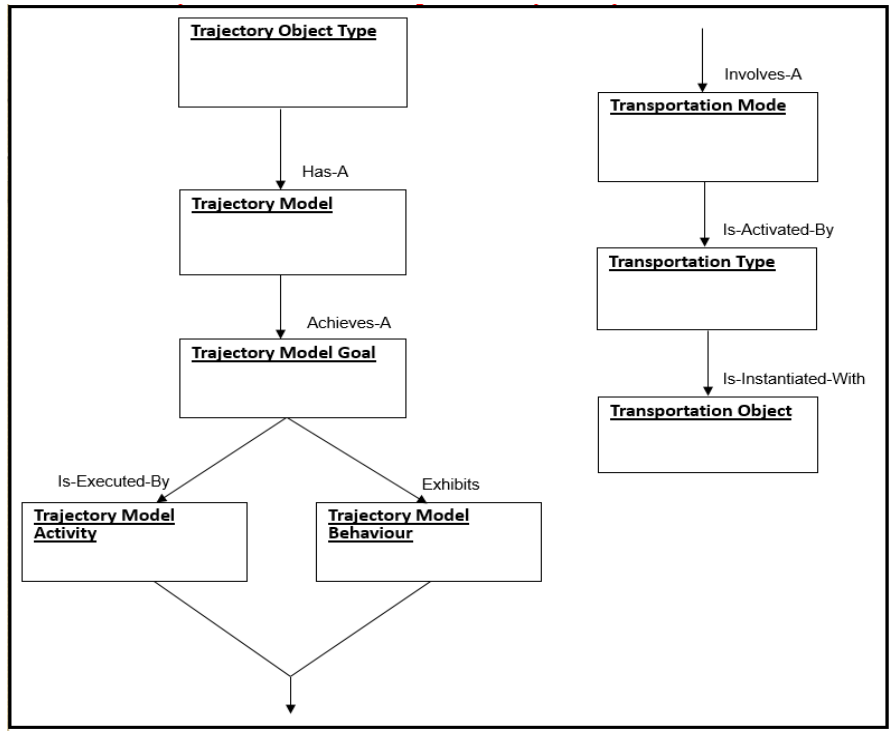

*Figure 9.*      *Taxonomy Model of the Trajectory Representation Dimension*

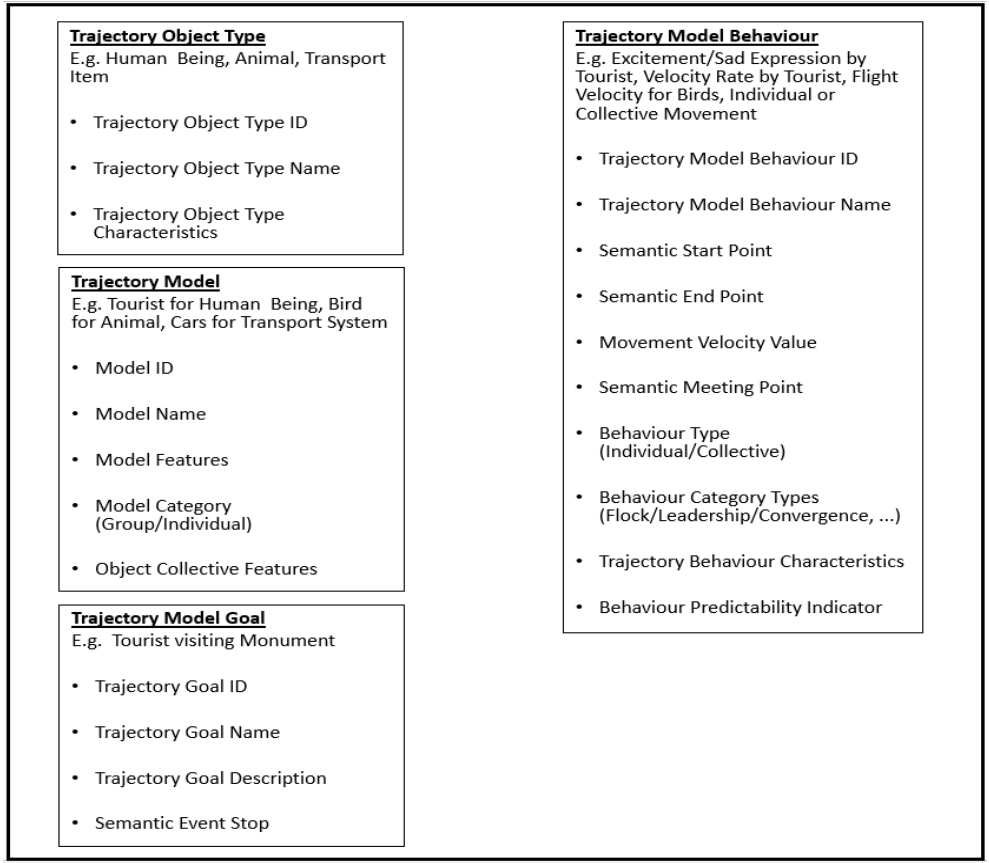

*Figure 10. Characteristic Description of Trajectory Representation Taxonomy Model (a)*

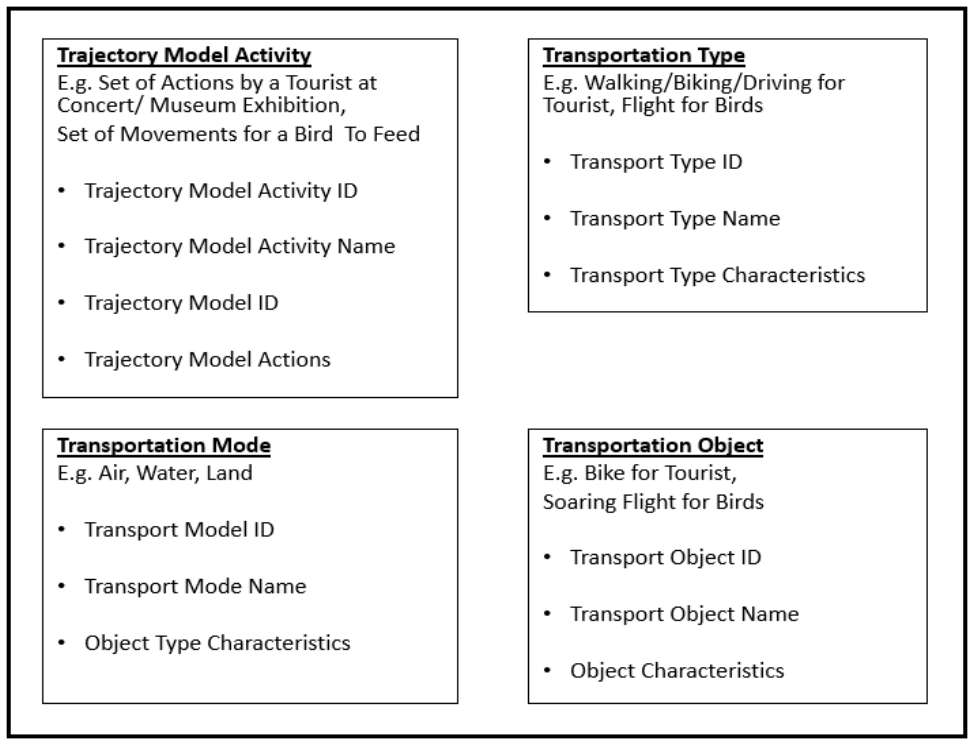

*Figure 11. Characteristic Description of Trajectory Representation Taxonomy Model (b)*

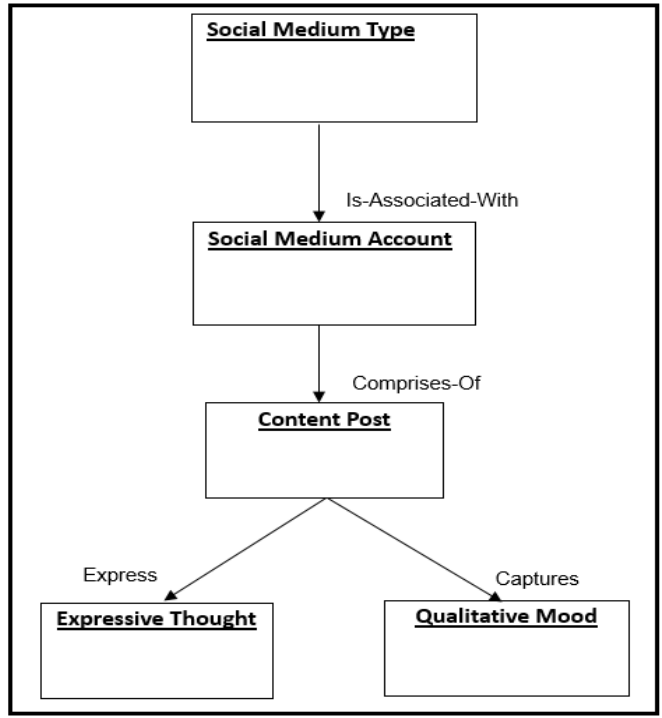

Figure 12. Taxonomy Model of the Social Interaction Dimension

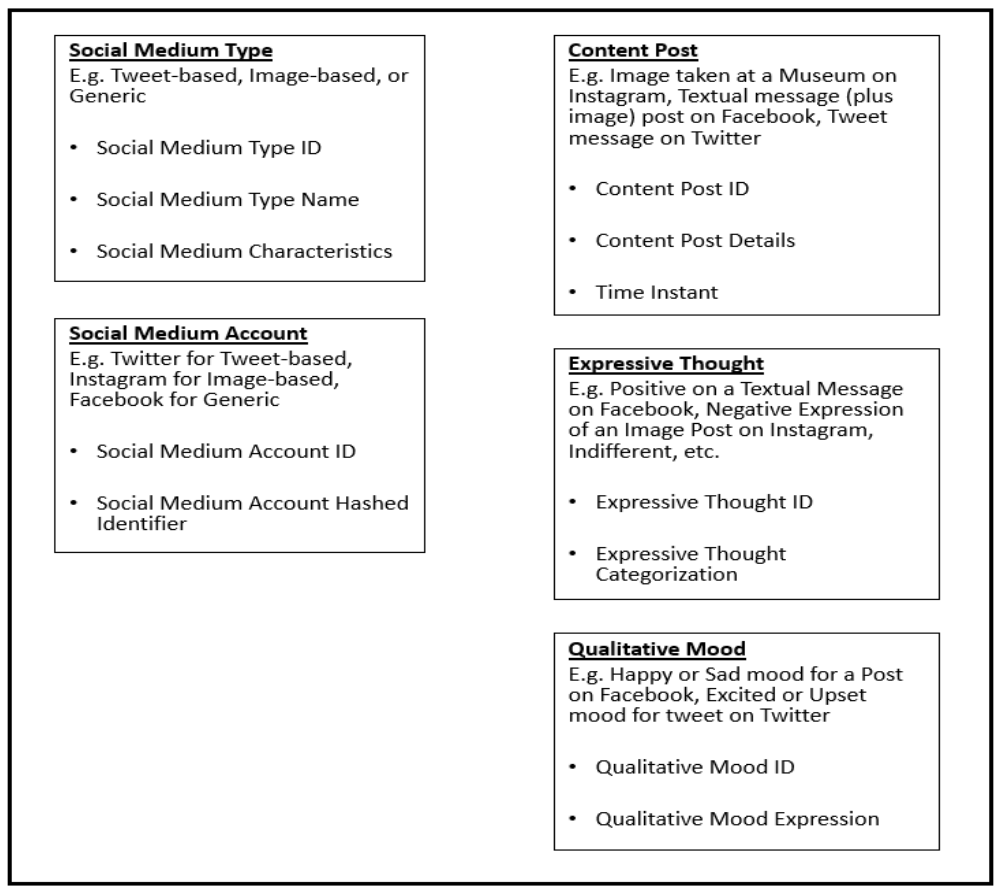

Figure 13. Characteristic Description of the Social Interaction Taxonomy Model

## Social Interaction Dimension

The Social Interaction thematic dimension captures and stores information on the trajectory object's interaction on social networks. This dimension captures data, such as the medium used to express the mood and thoughts, the comments or posts sent out, the qualitative mood, and the expressions, amongst others.

A representation of the social interaction dimension is modelled in the following hierarchical levels; namely, *Social Medium Type* (for e.g., tweet-based, picture-based, generic, etc.), *Social Medium Account* (for e.g., Facebook, Twitter, Instagram, Flickr, and Myspace, amongst others), *Content Post* (for e.g., textual message, and image, amongst others), *Expressive Thought* (for e.g., positive, negative, and indifferent, amongst others), and *Qualitative Mood* (for e.g., happy, sad, upset, and anxious, amongst others). *Figure 12* displays the taxonomy ontology model of the social interaction dimension depicting all the hierarchical levels expressed in the dimension.

It will be noted that procedures for hierarchy level aggregation and summarizations of the attribute data and characteristics modelled in this dimension will adapt and leverage on some of the methodologies crafted in social network data management and analysis in the literature. *Figure 13* illustrates an elaborate definition and characteristic outlook of the social interaction dimension which highlight the constituent attribute data item elements for each hierarchical level.

## EXPERIMENTAL IMPLEMENTATION APPROACH

This Section describes the procedural steps necessary in modelling and designing the generic semantic trajectory data warehouse. The key implementation procedures are Semantic Web Ontology Modelling, ETL Procedures, and Physical Trajectory Data Warehouse Design. The rest of the section explains some of the specific tasks that are involved in each of the procedures.

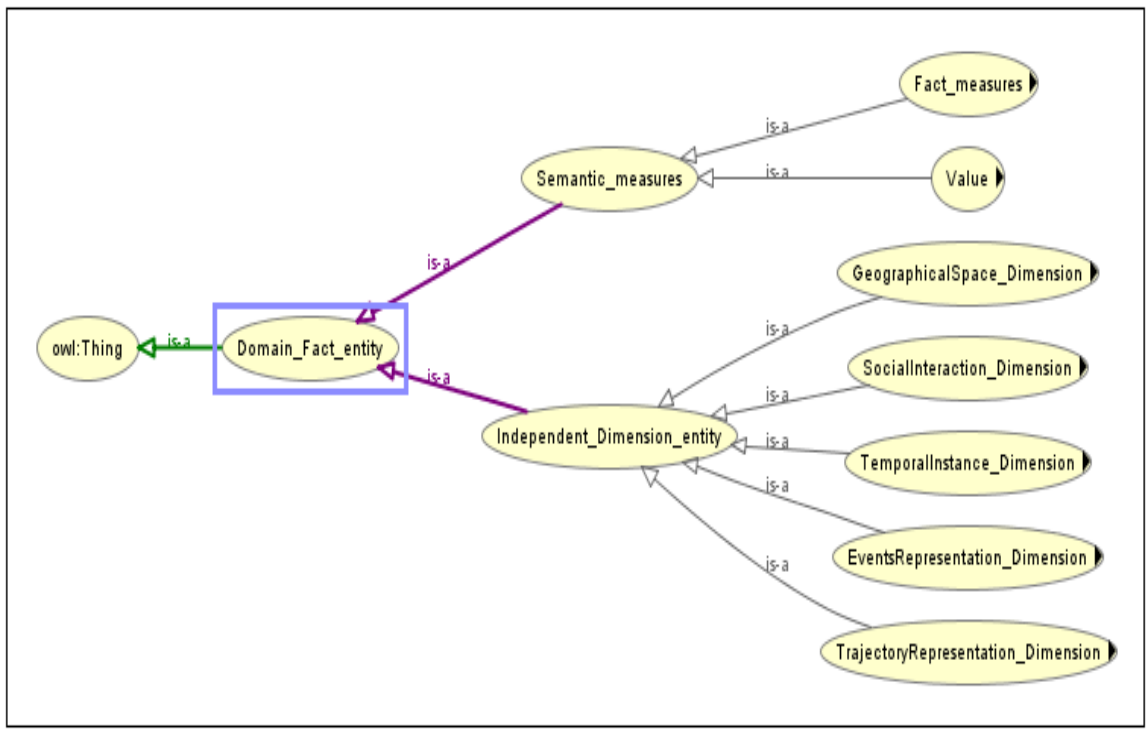

*Figure 14.    Web Ontology Language Modelling of the Semantic Trajectory DW*

## Semantic Web Ontology Modelling

The ontology modelling of the trajectory data warehouse is implemented using the framework architecture in Protégé Semantic Web Ontology platform. The ontology design approach is similar to prior work by the authors in (Manaa *et al.*, 2016; Sakouhi *et al.*, 2014; Yan *et al.*, 2008) and it is based on the formulations by W3C Web Ontology Language (OWL). OWL is a Semantic Web language designed to represent rich and complex knowledge about *things (subjects)*, groups of *things*, and relations between *things*.

To model the generic ontology, the general procedure is explained, as follows: a main class of a *Thing* (see *Figure 14*) is created. This class represents the universe of the domain. The *Domain_Fact_Entity* class is subsequently created, which serves as the main parent class for the application domain, and the only sub-class for the universal *Thing* class.

The general *Domain_Fact_Entity* class further contains two other sub-classes; namely, *Semantic_Measures* and *Independent_Dimension_Entity*. The *Independent_Dimension_Entity* sub-class represent the thematic dimension constructs which have been described in Section 4 and this class has 5 other sub-classes classified as *TemporalInstance_Dimension*, *GeographicalSpace_Dimension*, *EventsRepresentation_Dimension*, *TrajectoryRepresentation_Dimension*, and *SocialInteraction_Dimension*.

The *Semantic_Measures* sub-class is created with sub-classes of *Fact_Measures* and *Value*, of which the *Value* sub-class is defined with other sub-classes of *Object_Values* and *Data_Values*. The *Fact_Measures* models all the fact numeric measures defined for a particular trajectory movement, and is defined with sub-classes of *SquareArea*, *AverageEventTimeDuration*, and *MaxTrajectoryTravelDistance*, amongst others. The *Data_Values* sub-class models all the fact attribute data covering the trajectory movement, and it may be defined with sub-classes, such as, *Height*, *Depth*, *Width*, and *SemanticMeetingPoint*, amongst others. *Figure 14* illustrates the overview of the general ontology modelling of the semantic trajectory data warehouse. In *Figures 15* and *16*, the diagrams illustrate the OWL modelling constructs for the Fact attributes and measures, and the thematic dimension attributes and with hierarchical levels, respectively.

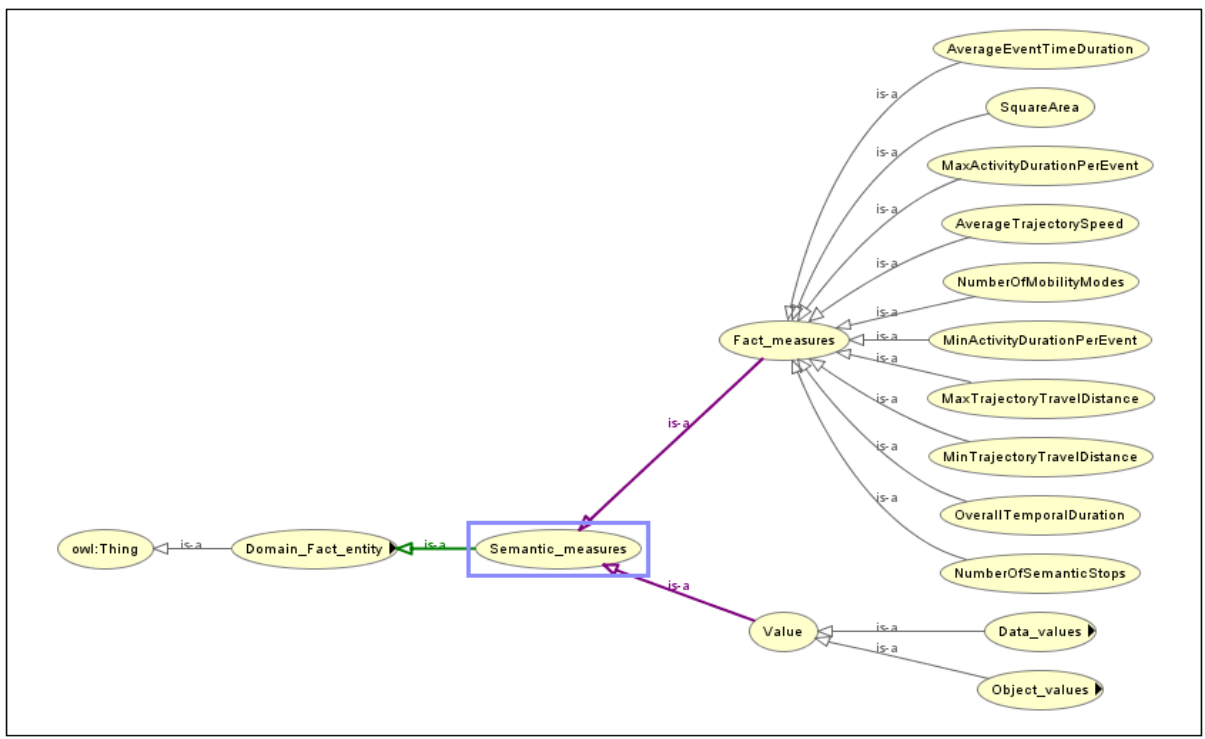

*Figure 15. OWL Modelling of the Semantic Trajectory DW – Fact Attribute & Measures*

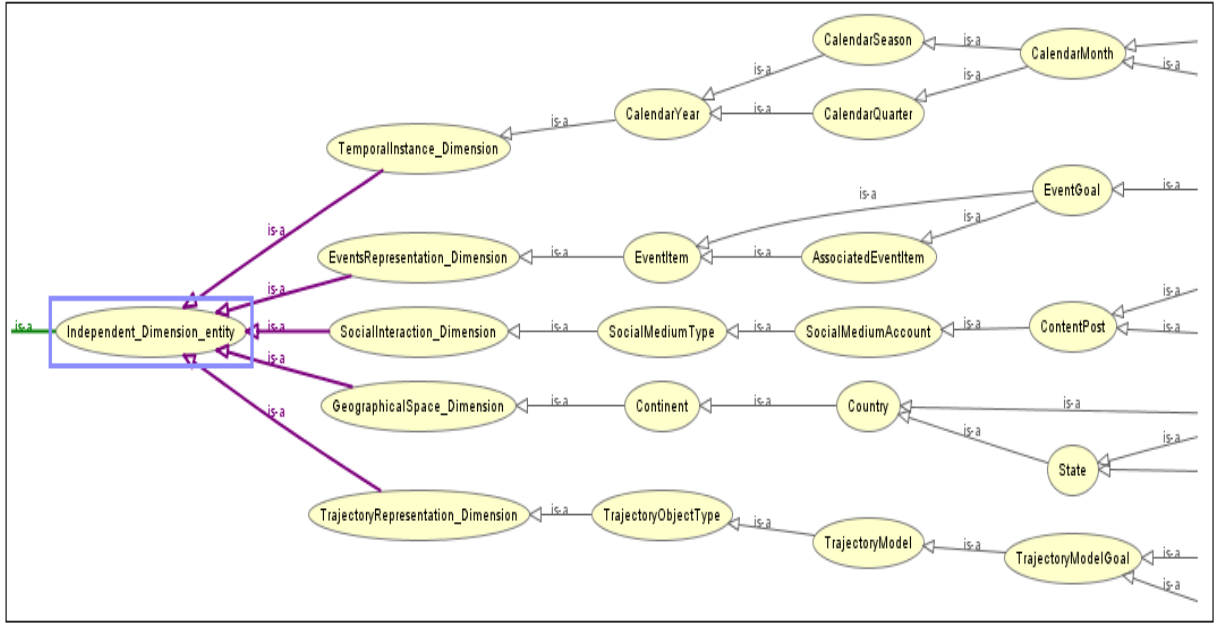

*Figure 16. OWL Modelling of the Semantic Trajectory DW – Dimensions*

## Extract-Transform-Load (ETL) Procedures

The ETL procedures are made up of tasks for the extraction, transformation, and loading of trajectory-related *raw data* into the data warehouse. The extraction procedures constitute a set of computing tasks and activities that help to retrieve heterogeneous data from varied sources to be transformed into a homogeneous data set. The design of the trajectory data warehouse is composed of a number of dimensions with different characteristics of data. As a result data values from geographic points, lines, and polygons have to be identified and extracted. Other forms of data are, the set of event data that the trajectory object participated in, the environmental data, as well as, social media data in relation to the trajectory dynamics. The *raw data* set extracted needs to be transformed into a format in which they can be suited as inputs into the fact and dimension table attributes in the data warehouse. Hence, data transformation tasks are performed consequently before the refined data is finally loaded into the warehouse data repository.

It will be noted that ETL procedures are not a core task in the overall data warehouse design and data population of the semantic trajectory data warehouse. Hence, the assumption is that a higher granularity of ETL-processed data is expected to be incorporated in the overall methodology. Consequently, a comprehensive evaluation and results analysis is realized.

Zekri and Akaichi (2014) in their recent study proposed a methodology approach for processing ETL procedures on trajectory data. In their approach, the authors propose a conceptual modelling of ETL processes and algorithms in order to implement trajectory ETL tasks. This methodology approach can be adopted for ETL implementation. Additionally, other traditional approaches can be adopted to perform the ETL procedures, and a number of application softwares exists which can be used. A typical example is the open-source application software framework of Pentaho Data Integration.

## Physical Design of Semantic Trajectory Data Warehouse

The physical design of the semantic trajectory data warehouse involves creating a data warehouse in a chosen Relational Database Management System (RDBMS). The constituent fact and dimension tables which make up the data warehouse are subsequently created with their domain attributes and data types. Each fact table is created with attributes and numeric measures and with a composite primary key being the foreign keys from each of the dimension tables. The dimension tables are created with their respective descriptive attributes and primary keys. For each dimension table, hierarchical levels needed for data aggregation and summarization are defined. This enables easy drill-up and roll-down of aggregate data during trend analysis and data visualization. Orlando *et al.* (2007) define practical approaches in defining the hierarchies and aggregation of spato-temporal data in a trajectory data warehouse.

A typical geospatial database management system that is adopted to implement the physical design of the semantic trajectory data warehouse. PostGIS (spatial and geographic extension of PostgreSQL) DBMS is adopted for the design and implementation of the semantic trajectory data warehouse. This application tool offers broad extension of spatial and geographic object features and implementation. Moreover, it offers an open-source license and development platform, and DBMS is incorporated with varied domain geographic data types. These functionalities and configurations make it easy to integrate as a back-end database server repository for data analytics softwares. *Figure 17* displays the physical database schema design for the semantic trajectory data warehouse. The illustration depicts the attribute domain information and the referential relationship for the fact and dimension tables.

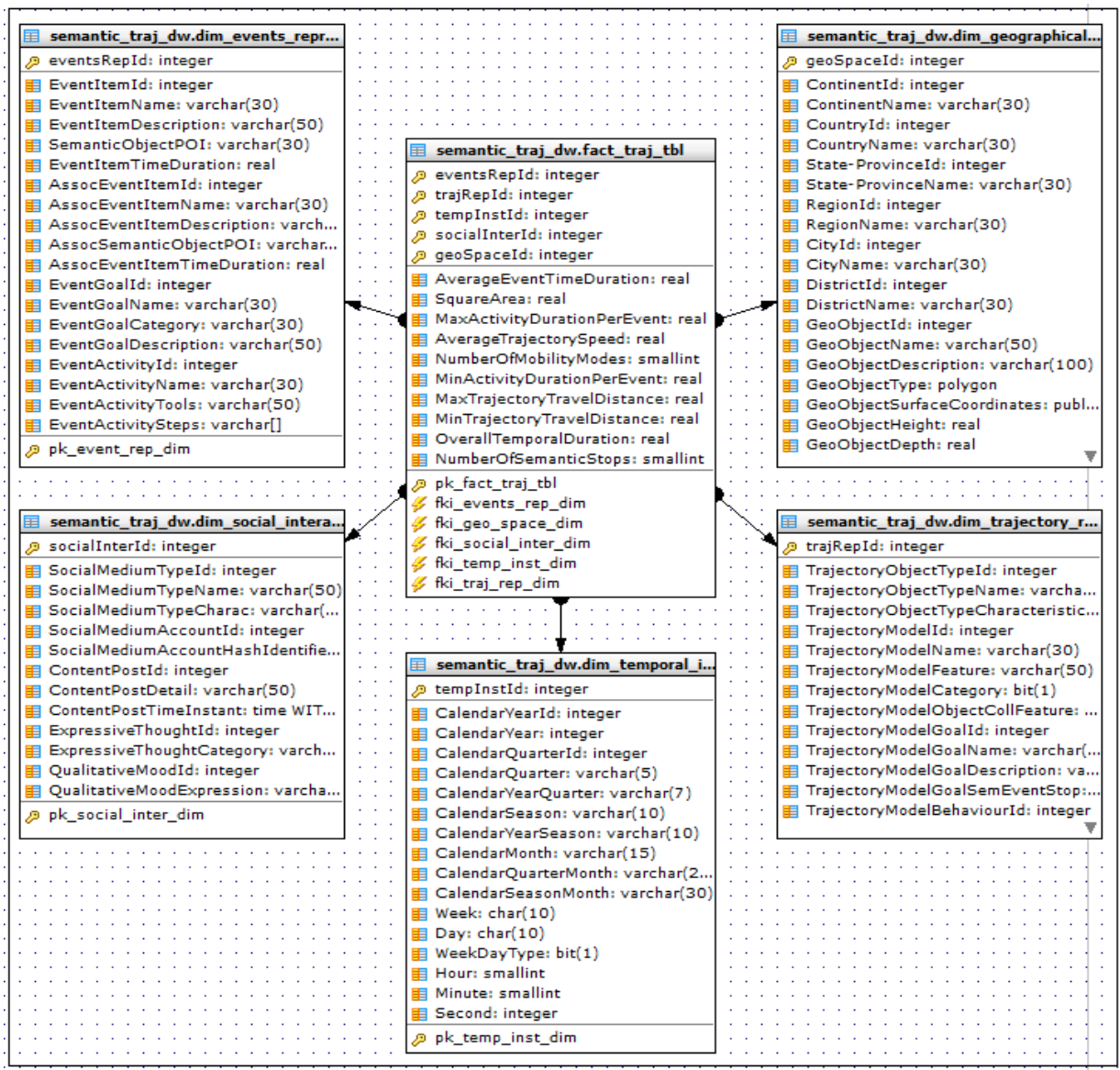

*Figure 17. Schema Design of the Semantic Trajectory Data Warehouse*

# APPLICATION DOMAIN EVALUATION & ANALYSIS

The study of trajectories and the pattern analysis of trajectory objects, associated events and activities, and semantics have been very useful in a number of domain applications. Most of these application studies have opened up some unheeded dynamics and better understanding of semantics regarding these domains. Additionally, the studies in these applications have offered critical solutions and innovations in prevalent, persistent problems; especially in the data processing activities on these domains.

This Section discusses 3 main application domains in which study of trajectories and semantic data warehouses have been applied and become useful in drawing out vital knowledge in data processing. These are namely, tourist movement and tourism management, animal (bird) migration and ecology, and urban traffic and transportation management, amongst others. Moreover, a breakdown analysis of the geographical domain, the trajectory objects, events and activities, as well as environmental factors for the study of trajectories on these application domains is performed. An evaluation for the expected results on the formulation and modelling of a generic ontology for trajectories, and the practical techniques in designing a semantic trajectory data warehouse is discussed.

## Expected Answerable Queries

This sub-section describes some of the practical queries and instantiations that inform the outcomes of the research on trajectory objects and trajectory data warehouses. Additionally, the predictive trend and behavioural analyses associated with the movements of trajectory objects which can be deduced from the data warehouse is addressed. It will be noted that once the semantic trajectory data warehouse is designed for an application-specific domain, a number of answerable queries can be posed to it.

To expatiate on the research results, evaluation queries that will illustrate the forms of semantic trajectory information expected is formulated for each form of application domain. These formulated queries investigate the data regarding the patterns and behaviours of the trajectory objects, the environmental objects, and the events, amongst others.

The formulated queries are contextualized under specific application domains to facilitate a better understanding of the expected results. Outlined in the following sub-sections are the application domains and their respective formulated instance queries.

### *Tourist Movement and Tourism Management*

The study of tourist behaviour and the management of tourism activities within a geographical area has also gained much attention by researchers. Tourism management generates quite an amount of data which needs to be processed and analyzed to identify peculiar patterns and even predict future trends. Additionally, the amount of revenue generated by tourism gives much concern for governments agencies and city authorities to plan and project various ways of sustaining and increasing revenue levels.

The trajectory patterns of tourists is inherent with raw data of which when these data values are harnessed, processed, ware-housed and mined, can offer vital information to the overall management of tourism. These vital information can help identify unique behaviours of different categories of tourists for different calendar seasons. Additionally, research into the trajectory patterns of tourists can help determine notable *Events-of-Interests* (events and activities; for example, carnival festivals, native festivals, etc.), and categories of tourists who patronize these events during specific temporal periods. Moreover, the association of trajectory movement pattern to *Points-of-Interests* (Landmark) objects and notable landmarks have also gained much interest in the literature and studies are on-going to ascertain these semantic associations.

Recent study by Bermingham and Lee (2014) studied and proposed methodologies in mining valuable data for the tourism industry in Queensland, Australia. In their study, the authors aimed at extracting valuable tourist information within a geographical region about where people go, at what time these tourist visit such places, and where the tourists are likely to go next. The authors adopted an approach of extracting spatio-temporal meta-data of tourist photos from the social media platform of Flickr.

Suppose we have an instance of a tourism management application domain for tourists in a city. The city has a lot of monumental places and sites, as well as landmark buildings and objects. Hence, in this case study we will like to investigate the trajectory movement patterns of a tourist taking into consideration the Point-of-Interests (POIs), events and relationship of the individual behavioural mode of touring.

**Query 1.** *What is the most popular type of Event occurrence {Festival, Carnival, Exhibition} for a travelling Tourist on a particular Purpose that occurs during a particular Calendar Season {Summer, Fall, Winter}? Why and which activities make the event popular and give the trends in Pattern and Average Time spent per a Tourist over the past 3 years.* ∎

## *Formulated Query 1:*

```
SELECT e.EventItemName, e.EventGoalName
FROM fact_traj_tbl f, dim_events_representation_tbl e,
dim_geographical_space_tbl g, dim_temporal_instance_tbl t,
dim_trajectory_representation r, dim_social_interaction s
WHERE f.eventsRepId = e.eventsRepId
AND f.trajRepId = r.trajRepId
AND f.tempInstId = t.tempInstId
AND f.geoSpaceId = g.geoSpaceId
AND f.socialInterId = s.socialInterId
AND t.CalendarSeason = 'Summer'
AND ST_WITHIN (g.GeoObjectType, ST_GeomFromText('POLYGON((-34.954449  -
8.124354, -34.904449 -8.124354, -34.904449 -8.084354, -34.954449 -8.084354,-
34.954449  -8.124354 ))', 4326))
AND COUNT (e.EventItemName) =
(SELECT MAX(COUNT(e2.EventItemName)) FROM dim_events_representation_tbl e2
GROUP BY e2.EventItemName)
GROUP BY e.EventItemName, e.EventGoalName
```

In *Query 1*, the formulated query wants to draw knowledge in the kind of *episodes* (segments of trajectories between subsequent stops) and occurrences of interesting activities for which a tourist participates. These episodes are usually associated with some Point-of-Interests (POIs). The semantic information of all these episodes and POIs are extracted to draw meaning on the trajectory and its movements. It will be noted that there is a temporal significance of a *Calendar Season*, as each season makes meaning for different trends of *Event* occurrences. To this end, the typical events that tourists are attracted to for a particular period, the time duration spent at these events, and what informs their attention to these events; as well as the impact it has on their overall trajectory patterns are drawn out.

**Query 2.** *What is the typical Profile of a Tourist for a particular Calendar Season {Winter, Summer, Spring}? What are the Relevant Stops, Tourist Behavioural and Transportation Modes and the Velocity Rate (movement speed) for the Tourist's movement.* ∎

In *Query 2*, the detailed analysis of the tourist's profile is extracted and analyzed. This form of profile analysis of the trajectory object could have associations with other objects in the spatio-temporal environment, and as a result infer on the trends of *Revelant Stops*, *Behavioural* and *Transportation Modes*, and the *Velocity Rate* for the movement. Moreover, the analysis of such profile is aware of the temporal period (*Calendar Season*) that the trajectory object (tourist) moves, as each period could exhibit different kinds of tourists and profiles.

*Formulated Query 2:*

```
SELECT r.TrajectoryModelName, r.TrajectoryModelBehaviourName,
r.TrajModelBehaviourMovementVelocity, r.TrajectoryTransportationModeName,
r.TrajectoryTransportationTypeName
FROM fact_traj_tbl f, dim_events_representation_tbl e,
dim_geographical_space_tbl g, dim_temporal_instance_tbl t,
dim_trajectory_representation r, dim_social_interaction s
WHERE f.eventsRepId = e.eventsRepId
AND f.trajRepId = r.trajRepId
AND f.tempInstId = t.tempInstId
AND f.geoSpaceId = g.geoSpaceId
AND f.socialInterId = s.socialInterId
AND t.CalendarSeason = 'Winter'
AND ST_WITHIN (g.GeoObjectType, ST_GeomFromText('POLYGON((-34.954449  -
8.124354, -34.904449 -8.124354, -34.904449 -8.084354, -34.954449 -8.084354,-
34.954449  -8.124354 ))', 4326))
```

## Birds Migration and Ecology

The study of the trajectory and migration patterns of animals in an ecology is an important subject for wildlife conservation and management. Much more importance is the ability to predict the sustainability of these animals in the ecology in the face of uncertain climate and food security. One kind of animal that has gained enough trajectory studies in the literature has been birds. Here, studies have focused on their migration in search food availability and suitable environment for breeding.

Oleinik *et al.* (2009) in their research studied about how environmental climatic changes affect the trajectory migration of birds. The research focused on *White Stork* breed of birds. These birds migrate from the Central and Western parts of Europe (Northern Hemisphere) to the Western, Central and Southern parts of Africa (Southern Hemisphere). The migration from Europe usually takes place in the Fall season, when the climate begins to change to colder and unfavourable periods for food availability. Subsequently, in the Spring season, the birds migrate back from the various parts of Africa to Europe for breeding and better food availability.

The key attention for trajectory studies and data warehousing on bird migration by Oleinik *et al.* (2009) and other researchers is the impact of each kind of climate change and the correlation, such as, the wind direction and speed, the temperature and humidity level, and the rain precipitation. Additionally, the prominent questions that were analyzed and will seek solutions are, as follows: What is the search space? What is the geographical region population concentration or density size? When do the birds arrive and set off?

Suppose there is the need to study and analyze the migration pattern of certain types of animals (for e.g., birds) within a specified period of time. In this paradigm, we will want to know what informs the flight of the birds at particular a temporal instance or season, the weather patterns that impact on their migration or movement, the food availability and sustainability, and the geographical surfaces of mountains, valleys, and hills, amongst others. These information tend to serve as valuable knowledge in the study on their sustainability in the ecology and migration patterns.

**Query 3.** *What is the average trajectory speed that particular characteristics of a Bird travels in between two known semantic stops? What is the Wind Direction, Temperature, and Rain Precipitation to force such pattern of speed between the stops. What are the trends over the past 2 years.* ∎

*Formulated Query 3:*

```
SELECT f.AverageTrajectorySpeed, r.TrajectoryModelName,
r.TrajectoryModelFeature, r.TrajModelBehaviourMovementVelocity,
e.EventEnvironmentType, e.EventEnvironmentCharac
FROM fact_traj_tbl f, dim_events_representation_tbl e,
dim_geographical_space_tbl g, dim_temporal_instance_tbl t,
dim_trajectory_representation r, dim_social_interaction s
WHERE f.eventsRepId = e.eventsRepId
AND f.trajRepId = r.trajRepId
AND f.tempInstId = t.tempInstId
AND f.geoSpaceId = g.geoSpaceId
AND f.socialInterId = s.socialInterId
AND ST_WITHIN (r.TrajSegmentSemanticStartPoint,
r.TrajSegmentSemanticEndPoint)
```

In *Query 3*, an indepth analysis of the characteristics of certain species of birds are analyzed in context. The research investigation here is to know the flight patterns and the kind of weather characteristics that impact on the speed of their flights over a specified temporal stance. These analyses tend to provide important research data for the wildlife and ecology management of these different species of animals.

*Highway Traffic and Transportation Management*

Transportation management involves the efficient usage of roads and highways and the minimization of medium to heavy traffics on the highways. Moreover, the need to ensure maximum safety of road users on the highways have become critical needs for city authorities and governmental agencies. In most cases, the varied usage of road networks gives rise to the planning and projections on the need to guarantee the control of speed or velocity rates, and the construction of road components (for e.g., interchanges, lane expansion, etc.) at certain parts of the highway.

The research on trajectories focusing on the application domain of highway traffic and overall transportation management has gathered much study in the literature more recently (Heo *et al.*, 2013; Jenhani *et al.*, 27). Some aspects of these research study investigates the various features and characteristics of the components of road parts, such as, the steepness of a road segment, and the events (activities) that occur at some points of the road network. The studies also examine the effect of these component characteristic features on safety of pedestrians and other road users, and the susceptibility and likelihood occurrence of collision accidents, amongst others.

It is of the general expectation that the data values on the trajectory movements of vehicular objects and their movement characteristics are collected and stored in a data warehouse. Afterwards, data mining procedures can be performed on the data repository to ascertain the pattern behaviour of the vehicular speeding, stopping, or curve negotiation. Moreover, the contributing factors to persistent traffics at certain sections of the road network can be identified and the problem solved, subsequently.

**Query 4.** *What segment of the Highway and what Hour Interval do most cars have an average speed less than 50 km/hr? What Events {Bridge Repair, Steep Slope, Street Carnival, Sharp Curves} and Activities affects the trends over the last 3 years.* ∎

*Query 4* illustrates a kind of research investigation to determine the typical movement patterns of vehicular objects that prevail on the highways. Here, there is the need to ascertain which parts of the road exhibit rare patterns of speed of moving cars, and if there are any notable events or associated activities that inform these changes in the speed patterns of the cars over a temporal instance. The information analyzed here offers vital information to city authorities to efficiently plan and project better road management practices, road safety, pollution control, and environmentally sustainable energy usage.

*Formulated Query 4:*

```
SELECT r.TrajSegmentSemanticStartPoint, r.TrajSegmentSemanticEndPoint,
f.AverageTrajectorySpeed, r.TrajectoryModelName, e.EventItemName,
e.EventActivityName
FROM fact_traj_tbl f, dim_events_representation_tbl e,
dim_geographical_space_tbl g, dim_temporal_instance_tbl t,
dim_trajectory_representation r, dim_social_interaction s
WHERE f.eventsRepId = e.eventsRepId
AND f.trajRepId = r.trajRepId
AND f.tempInstId = t.tempInstId
AND f.geoSpaceId = g.geoSpaceId
AND f.socialInterId = s.socialInterId
AND f.AverageTrajectorySpeed < 30
AND t.CalendarYear BETWEEN '2010' AND '2015'
```

*Table 1. Comparative Analysis of Application Domain Trajectory Dynamics*

| Trajectory Dynamics / Application Domain | Tourism Management | Birds Migration & Ecology | Highway Traffic & Transportation Management |
|---|---|---|---|
| 1. Trajectory Object | Human Being (e.g. Tourist) | Bird | Car, Family Vans, Truck |
| 2. Transportation Modes | Air, Water, Land | Air | Land |
| 3. Major POI Objects | Hotel, Castle, Museum | Mountain, Valleys, Tree | Highway Interchange, Bridges |
| 4. Transportation Types | Walking, Biking, Driving, Parachuting | Flight, Soaring | Driving |
| 5. Major Trajectory Goals | Entertainment at Concert, Scientific Interest at Museum | Food Availability, Environmental Conditions for Breeding | Speed Limit Observance, Monitor Accident Occurrences |
| 6. Major Trajectory Events | Museum Exhibition, Theater Concert | Feeding, Resting | Bridge Repair, Steep Slopes, Carnivals Festivities |
| 7. Major Trajectory Activities | Party, Musical Shows at Concert, Watching Movies | Sitting, walking in a nest on Mountain top, Consecutive picking of fruits, seeds, and insects with the beak | Negotiating Curves, Slow Acceleration and Interaction with Celebrator at Carnival road zones |
| 8. Environmental Factors | Cloud Overcast, Temperature, Rain Precipitation | Wind Direction, Temperature, Rain Precipitation | Snow Fall, Rain Precipitation, Fog Concentration |

## Comparative Analysis of the Trajectory Dynamics for Application Domains

In this Section, a comparative analysis of characteristics of trajectories to each of the application domains is presented. Here, the characteristics features covering the trajectory transportation modes, trajectory goals, major events and activities, and the environmental factors that affect the trajectory movement are outlined. *Table 1* above summarizes the trajectory dynamics for each of the application domains.

## COMPARISION & PERFORMANCE ANALYSIS TO OTHER APPROACHES

Earlier methodology approaches to the modelling of semantic trajectory data warehouses explain important aspects of varied semantic annotations, as well as the data item elements that define the semantics at each stage of the trajectory movement. On one hand, some methodologies focus on the appropriation of semantics to each dimensionality module of the data warehouse; by identifying and defining the application-specific characteristics that are associated with these modules. For example, the set of typical characteristics of a tourist, such as, the race, age band, income, and the specific reason on why he or she participates in particular events in its trajectory.

On the other hand, some of the methodologies address the concept of data warehousing using ontologies which are modelled to focus on dimensionality components of the data warehouse, such as, geographical and geometric modules. Moreover, some methodology approaches model the data warehouse according to specific application domains, such as, tourism management, and highway transportation management, amongst others.

This Section compares and discusses the performance analysis on the characteristic propositions outlined in prior background work in relation to the novel methodology propositions described in this research (in Section 4). The comparison discussion is focused on two recent and prominent methodologies in the literature; namely, Da Silva *et al.* (2015) and Manaa and Akaichi (2016). *Table 2* addresses and outlines the comparative analysis based on the following criteria; methodology approach, social interaction inferences, domain applications, level of semantic enrichment, operational scalability, and level of practical query processing and optimization, amongst others.

## Discussions on Qualitative Comparative Analysis and Performance Measures

It will be accessed from the above comparative analysis that the methodology approach presented in this project research offers a better platform for modelling of generic semantic trajectory data warehouse. A key project assumption that has to be highlighted in terms of the comparative analysis is that, the research project was not experimented in terms of populating the data warehouse with ETL-processed data and the associated practical query processing.

The major comparison and performance analyses in *Table 2* above is discussed. In terms of methodology approach, the proposed methodology offers a complete ontology modelling that addresses every unique facet for any application domain expected, in comparison to the propositions in the literatures Da Silva *et al.* (2015) and Manaa and Akaichi (2016). Manaa and Akaichi (2016) present an ontology approach but it is not comprehensive enough to incorporate most semantic annotations for trajectories, as well as giving a practical approach. On the other hand, Da Silva *et al.* (2015) present an ontology approach presented in conceptual layers and does not outline a design outlook as to how to practically implement the propositions.

With regards to adding enough social interaction semantic annotations, it will be inferred that the proposed methodology provides the social media data into the trajectory data warehouse, whereas the other two methodologies do not offer such semantic data. This is a major input in enriching the semantic data annotations for the data warehouse.

*Table 2. Qualitative Methodology Comparative Analysis and Performance Measurements*

| Criteria/Methodology Approach | Da Silva *et al.* (2015) | Manaa & Akaichi (2016) | Proposed Methodology (2017) |
|---|---|---|---|
| 1. Methodology Approach Proposition | The methodology formalizes a conceptual approach in designing semantic trajectory data warehouse that relies on the DOGMA framework (Jarrar *et al.*, 2008). The approach offers a dual modelling of ontologies into two conceptual data layers, namely; *Consensual* and *Interpretation*. | The methodology adopts an ontological approach to model the constructs of the trajectory data warehouse using Ontology-Based Moving Object Data (OBMOD) to query heterogeneous data. The authors formulated an algorithm procedure to design the trajectory data warehouse schema. | The methodology adopts an ontological approach to the modelling of the data warehouse. The approach uses a Multidimensional Entity Relationship (MER) orthogonal notation for the generic ontology and a star-schema model to design the semantic trajectory data warehouse. |
| 2. Operational Scalability | There is no clear discussion for an operational design and experimental implementation of a trajectory data warehouse. Hence, the approach does not offer an assessment for scalability for large trajectory data. | There is no assessment of operational scalability for trajectory data. The authors did not define practical design approaches or steps for a typical data population for the data warehouse. | The methodology design of the trajectory data warehouse proposes to offer a platform to store and process scalable data, better than previous approaches of Da Silva *et al.* (2015) and Manaa & Akaichi (2016), though full experimental implementation is yet to be completed. |
| 3. Social Interaction Semantic Data | Methodology approach does not incorporate data modelling on social media interaction. | Methodology approach does not incorporate data modelling on social media interaction. | Methodology approach adopts a taxonomy modelling on social media interaction; that adds additional semantic annotation to the overall modelling of the trajectories and the data warehouse repository. |
| 4. Domain Application | The methodology approach formalizes modelling constructs and categorizes contextual information (based on 6 perspectives of *who*, *what*, *when*, *where*, *why*, and *how*) that can be related to each application domain. | The approach outlines a modelling concept applicable to various application domains, better than the approach in Da Silva *et al.* (2015) (which does not adopt an ontology). This methodology approach makes it better suited for specific characteristics of the chosen application domain. | The modelling concept presents a complete generalized approach applicable to all kinds of application domains, their characteristics and semantics. The clearly defined ontology modelling constructs makes this approach better in analyzing the application domains than propositions in Da Silva *et al.* (2015) and Manaa & Akaichi (2016). |
| 5. Query Processing and Optimization | The methodology does not clearly define the design procedures for a trajectory data warehouse. Moreover, the approach does not offer practical methods of query processing and optimization of processed queries on the proposed trajectory data warehouse. | The methodology approach discusses optimization issues in the design of the trajectory data warehouse, but does explain indepthly the procedures to achieve good query processing optimization. | The methodology approach proposes to offer a better query processing and optimization platform, with a final completion of the comprehensive data population and query processing. |
| 6. Level of Semantic Enrichment | The approach outlines and offers appreciable levels of semantic information based on the 6 perspectives of contextual information formalized in the *Interpretation Layer* of the proposed SWOT model. | The approach offers appreciable levels of semantic information to the ontology modules defined in the geometric trajectory, geographic, and application domains. Moreover, the approach outlines the annotations on semantic Region of Interest (ROI) and goals behind the activities in these ROIs. | The approach incorporates a lot of semantic annotations to enrich the modelling constructs, through analyzing the reasons (goals or purposes) and trajectory patterns in the geographical space, events and its activity associations, trajectory object and its path, as well as, social media interaction. |
| 7. Granularity of Processed ETL Data for Storage in the Data Warehouse | The methodology approach does not specify the granularity level of processed ETL data. The authors state the incorporation of semantically enriched trajectory data as minimum granularity. | The methodology approach does not specify the granularity level of ETL processed data. The methodology was focused on ontology modelling for the trajectory data warehouse. | The methodology approach uses higher granularity of processed ETL data for storage in the physical database. This level of granularity is expected because of the unique definition of dimension attributes, and definite attribute and highly aggregated measure data in the fact table. |

On query processing and optimization, it is the expectation that the proposed methodology offers a better performance in comparison to the other approaches. Here, the comprehensive physical fact and dimension attribute description gives a complete metric for query processing and optimization measures. Star-schema model for data warehouses are noted for fast query processing. Hence, the adoption of a star-schema model for the trajectory data warehouse will offer a platform for higher query processing rate.

Finally, on the level of semantic annotation enrichment, the proposed methodology offers a higher level of semantic information to the underlying trajectory data for the data warehouse. The expressive definition of the thematic dimensionality constructs enables the incorporation of all relevant semantic data per the POI objects, events, associated events, unique activities, trajectory objects, and the geographical space and environment, amongst others.

## CONCLUSION

This paper presented a novel methodology approach for the modelling of generic ontology for trajectories. The methodology approach used the formulated ontology model to serve as a framework model for the modelling and design of a semantic data warehouse for trajectory objects in a spatio-temporal paradigm. The researcher addressed the taxonomy modelling and thematic constructs for the generic ontology; which are namely, geographical space, temporal instance, events representation, trajectory representation, and social interaction.

Moreover, the researcher discussed and analyzed the adoption of the conventional Multidimensional Entity Relationship (MER) notation for spatio-temporal data warehouse modelling and design. The design of the data warehouse constituted fact and dimension tables, with the fact table displaying an *n-ary* relationship to each of the dimension tables. The generic ontology model was implemented on Protégé Semantic Web Ontology Language framework software and the data warehouse was implemented on PostGIS object-relational DBMS. As part of a partial evaluation, the researcher discussed the key research outcomes or results for the project, and analyzed some formulated queries as instantiations of the research outcomes. These queries were analyzed under 3 application domains to highlight on the contextual information regarding the queries.

The proposed methodology approach in this research in relation to two recent methodology approaches was compared, and the merits that this novel approach offers over the others in the trajectory data processing was subsequently discussed. Finally, the researcher discussed some of the application domains that semantic trajectory data warehouses can be very relevant in the later part of this literature. In summary, the methodology approach presented offers domain experts, researchers, and practitioners with a framework model for modelling generic ontologies. Moreover, the methodology approach offers modelling and design criteria on efficient approaches to design a semantic trajectory data warehouse for any application domain.

**Open Issues:** As part of this research study on semantic trajectory data warehouses some areas of open issues have arisen. A typical issue of privacy has to be critically addressed when extracting relevant information from trajectory objects, such as, cars and human beings; and their associated POIs objects. This was highlighted by Parent *et al.* (2013) in their seminal paper on modelling and analysis of semantic trajectories. Additionally, in the context of extracting social media data, the privacy of the online account information has to be preserved (Bermingham *et al.*, 2014; Seidl *et al.*, 2016). To this end, the methodology has to adopt some privacy-preservation requirements and integrate practical measures of protecting the privacy of trajectory objects, especially the social media data on human tourists in a tourism application domain.

**Future Work:** A number of future research directions still remain. The ability to design the scalable data warehouse to handle large sets of trajectory data for an increasing volume of data that will be collected, processed, and stored. Moreover, the need to incorporate comprehensive optimization measures for faster and more efficient query processing should be addressed. Finally, the need to formulate and enforce a privacy policy framework on the modelling and design of the semantic trajectory data warehouse needs to be considered in future works.

# ACKNOWLEDGMENT

The researcher extends special gratitude to Valéria Cesário Times (at the Center for Informatics, Federal University of Pernambuco) for her profound and insightful thoughts in this research.# REFERENCES